\documentclass[aps,preprint,superscriptaddress,amsmath,amssymb,showpacs]{revtex4}
\usepackage{graphicx}
\usepackage{dcolumn}
\usepackage{bm}

\graphicspath{{./fig/}}  

\begin{document}


\title{Use  and Abuse of a Fractional Fokker-Planck
Dynamics for Time-Dependent Driving }


\author{E. Heinsalu}
  \affiliation{Institute of Theoretical Physics, Tartu University,
  T\"ahe 4, 51010 Tartu, Estonia}
  \affiliation{Institut f\"ur Physik,
  Universit\"at Augsburg,
  Universit\"atsstr. 1,
  D-86135 Augsburg, Germany}

\author{M. Patriarca}
  \affiliation{Institute of Theoretical Physics,
  Tartu University, T\"ahe 4, 51010 Tartu, Estonia}
  \affiliation{Institut f\"ur Physik,
  Universit\"at Augsburg,
  Universit\"atsstr. 1,
  D-86135 Augsburg, Germany}

\author{I. Goychuk}
  \affiliation{Institut f\"ur Physik,
  Universit\"at Augsburg,
  Universit\"atsstr. 1,
  D-86135 Augsburg, Germany}

\author{P. H\"anggi}
  \affiliation{Institut f\"ur Physik,
  Universit\"at Augsburg,
  Universit\"atsstr. 1,
  D-86135 Augsburg, Germany}

\date{\today}

\begin{abstract}

We investigate a subdiffusive, fractional Fokker-Planck dynamics
occurring in time-varying potential landscapes
and thereby disclose the failure of the fractional Fokker-Planck
equation (FFPE) in its commonly used form when generalized in an
{\it ad hoc} manner to time-dependent forces. A modified FFPE
(MFFPE) is rigorously derived, being valid for a family of
dichotomously alternating force-fields. This MFFPE is numerically
validated for a rectangular time-dependent force with zero average
bias. For this case subdiffusion is shown to become enhanced as
compared to the force free case. We question, however, the existence
of any physically valid FFPE for arbitrary varying time-dependent
fields that differ from this dichotomous varying family.

\end{abstract}

\pacs{05.40.-a, 05.40.Fb, 05.60.-k, 02.50.Ey}


\maketitle


Normal Brownian motion occurring on potential landscapes that vary
in time is known to exhibit a multifaceted  collection of
interesting phenomena, such as Brownian motors, anomalous nonlinear
response behaviors, and stochastic resonance \cite{Phen}, to name a
few. Therefore, it is tempting to ask, whether an explicit
time-dependent force entails a similarly versatile scenario also in
the case of anomalously slow relaxation processes, relevant in many
systems, such as polymer chains,  networks, proteins, glasses and
charge-carriers in semiconductors \cite{Scher}. This issue is in
fact contained already in the first works on the motion of
charge-carriers in semiconductors \cite{ScherMontroll}, and has been
the subject of some further investigations ever since, see e.g. the
works \cite{sokolov2001, Barbi, Sokolov2006, SokolovKlafter06}, but
never really has attracted proper attention on its fundamental
level. Ultraslow relaxation in time-dependent external
potential-fields thus still constitutes  a challenge that is far
from trivial.

A widely used approach to study subdiffusive processes is based on
the fractional Fokker-Planck equation (FFPE) \cite{metzler2000R,
letter},
\begin{eqnarray} \label{FFPE}
\frac{ \partial }{\partial t} P(x,t)=
\sideset{_0}{_t}{\mathop{\hat D}^{1-\alpha}} \left [ -
\frac{\partial}{\partial x} \frac{F(x)}{\eta_\alpha} + \kappa
_\alpha \frac{\partial ^2}{\partial x^2} \right ] P(x, t) \, .
\end{eqnarray}
Here $F(x)$ is the force, $\eta_\alpha$ is the fractional friction
coefficient, $\kappa_\alpha$ is the fractional free diffusion
coefficient, and $\sideset{_0}{_t}{\mathop{\hat D}^{1-\alpha}}$
denotes the Riemann-Liouville fractional derivative,
\begin{equation} \label{RL}
\sideset{_0}{_t}{\mathop{\hat D}^{1-\alpha}} \chi (t) =\frac{1}{
\Gamma(\alpha)} \frac{\partial}{\partial t} \int_{0}^{t}
\mathrm{d} t' \, \frac{\chi(t')}{(t-t')^{1-\alpha}} \, .
\end{equation}
For time-independent forces the FFPE (\ref{FFPE}) can be rigorously
derived from  continuous time random walk (CTRW) theory
\cite{metzler2000R, letter}, which corresponds to the
simple random walk, 
including the non-Markovian character of the process via a
Mittag-Leffler residence time distribution (RTD) $\psi (\tau)
\propto \tau ^{-1 -\alpha}$ ($0< \alpha <1$) \cite{Scher,
ScherMontroll, shlesinger1974}.

With this work we show that the FFPE (\ref{FFPE}) fails in the case
of a time-dependent force $F(x,t)$. Furthermore, we argue that the
FFPE (\ref{FFPE}), when generalized ad hoc to a time-dependent
force, does not correspond to a physical stochastic process. This
affects a still steadily growing body of current research
\cite{Sample}, and implies that these so obtained  results  therein
are physically defeasible.

In different context, the study of a subdiffusive dynamics  in the
case of a purely time-dependent force $F(t)$ has given rise to a
fractional Fokker-Planck equation which differs from
Eq.~(\ref{FFPE}) \cite{SokolovKlafter06}. In this letter, we derive
an equation of similar form for the class of dichotomously
alternating force-fields $F(x, t)=F(x)\xi(t)$ with $\xi(t)=\pm 1$,
varying in space and time. In the case of a Mittag-Leffler RTD it
reads,
\begin{eqnarray}\label{FFPEmod}
\frac{ \partial }{\partial t} P(x,t)= \left [ -
\frac{\partial}{\partial x} \frac{F(x,t)}{\eta_\alpha} + \kappa
_\alpha \frac{\partial ^2}{\partial x^2} \right ]
\sideset{_0}{_t}{\mathop{\hat D}^{1-\alpha}} P(x, t) \, .
\end{eqnarray}
Below, we prove this form in terms of CTRW theory  and additionally
validate its correctness via the comparison of the analytical
solutions of this so modified fractional Fokker-Planck equation
(MFFPE) (\ref{FFPEmod}) for a rectangular time-varying periodic
force $F(x, t)\equiv F(t)= \xi (t) F_0$
with the numerical simulations of the underlying CTRW. Different
force-fields, such as $F(x,t)=\pm \sin(x)$ have also been
successfully tested (the details will be presented in a longer
follow up work). Our main point is, however, that the reasoning
 provided in proving (\ref{FFPEmod}) forces us to scrutinize the
 {\it physical validity} of this so modified FFPE in~(\ref{FFPEmod})
 already beyond a dichotomous
driving $F(t)$; as e.g. it happens for a sinusoidal driving $\xi(t)$
as used in Ref. \cite{SokolovKlafter06}.

As an interesting result, we also show that a symmetric dichotomous
force with average zero bias enhances the diffusion in respect to
the free case. Furthermore, it is found that for sufficiently slow
driving the effective fractional diffusion coefficient
$\kappa_\alpha^{\mathrm{(eff)}}$ exhibits a maximum {\it vs.} the
fractional exponent $\alpha$.


It is well known that neither a non-Markovian Fokker-Planck equation
nor its solution with the initial condition $P(x,t)=\delta(x-x_0)$
(i.e. the two-event conditional probability $P(x,t|x_0)$) can fully
define the non-Markovian stochastic process \cite{HT}. This is due
to the fact that all non-Markovian processes, such as a CTRW with a
non-exponential waiting time distribution, lack  per definition the
factorization property, which would allow to express all the
higher-order (multi-event) probability density functions in terms of
the first two ones. Because a CTRW is at the root of the FFPE
 (\ref{FFPE}) \cite{metzler2000R, letter}, in order to
generalize the latter to the time-dependent forces, one again starts
from  CTRW theory. However, the usual scheme of merely replacing a
time-independent force $F$ in Eq. (\ref{FFPE}) in an {\it ad hoc}
manner with a time-dependent $F(t)$ is doomed to failure. The reason
is that the underlying CTRW possesses a RTD with an {\it infinite}
mean. Thus, any regular driving with a large but finite period is
nonadiabatic. This very circumstance lies at the heart of the
overall failure of Eq.~(\ref{FFPE}) for time-dependent force fields.

In terms of a renewal description, a CTRW is a  semi-Markovian
process, meaning that the sojourn times spent on the localization sites are
independently distributed. Let us consider a one-dimensional CTRW on
a lattice $x_i=i\Delta x$ ($i=0, \pm 1, \pm 2, \dots$). After a time
$\tau $ drawn from the RTD $\psi_i(\tau)$, the particle at site $i$
jumps with the probability $q_i^\pm$ to one of the nearest neighbor
sites. The external force-field $F(x)$ specifies both $\psi_i(\tau)$
and $q_i^{\pm}$, see Ref. \cite{letter}. Modulating the force $F(x)$
in time, $q_i^{\pm}$ assume obviously a time-dependence and
$\psi_i^{\pm}(\tau|t)=q_i^{\pm}(t+\tau)\psi_i(t+\tau,t)$ become
conditioned on the entrance time $t$ for the site $i$ \cite{PRE04}.
For a Markovian CTRW with time-dependent rates $w_i^{\pm}(t)$ it is
known  that
\begin{eqnarray} \label{change}
\psi_i(t+\tau,t)=w_i(t+\tau)\exp
\left[-\int_{t}^{t+\tau}w_i(t')dt'\right] \, ,
\end{eqnarray}
with $w_i(t)=w_i^{+}(t)+w_i^{-}(t)$ and
$q_i^{\pm}(t)=w_i^{\pm}(t)/ w_i(t)$.
 For a driven non-Markovian CTRW,
however, a relation similar to Eq.~(\ref{change}) is lacking. As a
result, the use of a FFPE when generalized to the time-dependent
case of a time-varying force-field remains moot. The usual scheme of
the derivation of the generalized FFPE from the underlying CTRW can
be used only if $\psi_i(\tau)$ remains {\it unmodified} by the
time-dependent fields, i.e. if only the jump probabilities $q_i^\pm
(t)$ change. We consequently find that
$\psi_i^{\pm}(\tau|t)=q_i^{\pm}(t+\tau)\psi_i(\tau)$. Thus, the RTD
$\psi_i(\tau)$ remains unaffected only in the case of a dichotomous
flashing force $F(x,t)=F(x)\xi(t)$, where $\xi(t)=\pm 1$ is a
general dichotomic function of time $t$ which can change
periodically or also stochastically. Then, $q_i^{\pm}(t)=
\exp(F(x_i)\xi(t)\Delta x/2)/[\exp(F(x_i)\Delta x/2)+
\exp(-F(x_i)\Delta x/2)]$. We assume that $F(x)$ is continuous.
Then, the MFFPE~(\ref{FFPEmod}) can be derived rigorously in the
continuous space limit. The derivation precisely follows the same
reasoning as detailed in Ref.~\cite{letter}, while taking
$\psi_i(\tau)$ as being the Mittag-Leffler distribution.
It must be emphasized that for other driving forms $\xi(t)$, e.g.
for a sinusoidal driving $F_0 \sin(\omega t)$, this outlined
derivation becomes flawed because $\psi_i(\tau)$ is affected by such
time-varying fields, as being unveiled already with
Eq.~(\ref{change}). We remark also that due to the weak ergodicity
breaking~\cite{BelBarkai, heinsalu2006b} this MFFPE (\ref{FFPEmod})
describes the dynamics of an ensemble of particles rather than the
dynamics of an individual particle.

Notably, Eq.~(\ref{FFPE}) applied to the case of a time-dependent
force may well define an interesting mathematical object in its own
right; its connection to a known physical process, however, remains
open to question. One may attempt to justify Eq.~(\ref{FFPE}) for a
time-inhomogeneous situation by appealing to the concept of
''subordination''; noting that this equation corresponds to a random
process described by a usual Langevin equation but with the
operational time being a random stochastic process
\cite{Stanislavsky, Magdziartz}. A time-dependent physical force,
however, varies in {\it deterministic},  real time, which physically
cannot be transformed to random time.

\begin{figure}[t]
\centering
\includegraphics[width=8.5cm]{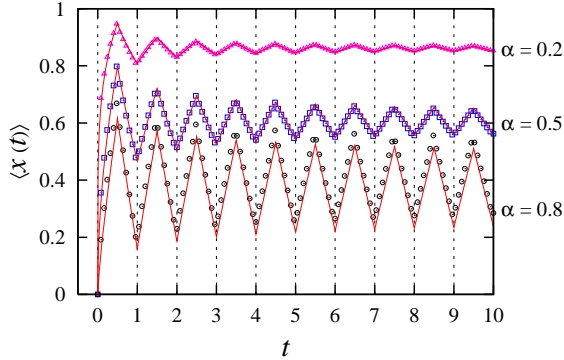}
\caption{(Color online) Average particle position $\langle x(t)
\rangle$ for various values of the fractional exponent $\alpha$:
Symbols represent the numerical results for the CTRW obtained by
averaging over $10^6$ trajectories, while continuous lines represent
the analytical solution (\ref{xx}) of the MFFPE~(\ref{FFPEmod}). The
time-period of the force is $\tau_0 = 1$ and
$F_0/(\eta_{\alpha}\sqrt{\kappa_{\alpha}})=1$ is used in numerical
simulations. The simulation algorithm is described in
Ref.~\cite{heinsalu2006b}. } \label{Fig1}
\end{figure}

In the following we study the particular case of a periodic
rectangular driving force $F(t)=F_0(-1)^{[2t/\tau_0]}$, where
$\tau_0$ denotes the time-period and $[a]$ is the integer part of
$a$. Put differently, we consider a dichotomous modulation of a
biased free subdiffusion where the absolute value of the bias is
fixed, but the direction of the force flips periodically in time.
The average bias is zero. Let us begin by finding the recurrence
relation for the moments $\langle x^n(t) \rangle$. Multiplying both
sides of Eq.~(\ref{FFPEmod}) by $x^n$ and integrating over the
$x$-coordinate one obtains,
\begin{eqnarray} \label{a2}
\frac{d\langle x^n(t) \rangle}{d t} &=& n F(t) D^{1-\alpha}_t
\langle x^{n-1}(t) \rangle/\eta_{\alpha}
\nonumber \\
&+& n (n - 1) \kappa_\alpha D^{1-\alpha}_t \langle x^{n-2}(t)
\rangle \, ,
\end{eqnarray}
for $n \ge 2$. When $n=1$ the last term on the right hand side of
the latter equation is absent. Then,
\begin{eqnarray} \label{xav}
\frac{d \langle x(t) \rangle}{d t} = \frac{F(t)}{\eta_\alpha}
D^{1-\alpha}_t 1 = \frac{F(t)}{\eta_\alpha \, \Gamma(\alpha)} \,
t^{\alpha-1} \, .
\end{eqnarray}
Integrating Eq.~(\ref{xav}) in time, the solution for $\langle x(t)
\rangle$ reads:
\begin{eqnarray} \label{xx}
&& \langle x(t) \rangle \nonumber \\
&& =\left\{
\begin{array}{c@{\quad \quad}}
x_N + \frac{v_{\alpha} t^\alpha}{\Gamma(\alpha+1)} \, , \quad
N\tau_0 \le t < (N+\frac{1}{2})\tau_0 \, , \\
x_N' - \frac{v_{\alpha} t^\alpha}{\Gamma(\alpha+1)} \, , \quad
(N+\frac{1}{2})\tau_0 \le t < (N+1)\tau_0 \, ,
\end{array}
\right.
\end{eqnarray}
where
\begin{eqnarray} \label{xn1}
x_N &=& \langle x(0) \rangle - \frac{v_{\alpha}
(N\tau_0)^\alpha}{\Gamma(\alpha+1)} \\
&+& \frac{v_{\alpha} \tau_0^\alpha}{\Gamma(\alpha+1)}
\sum_{n=0}^{N-1}\left[2(n+1/2)^\alpha-n^\alpha-(n+1)^\alpha\right]
\, ,
\nonumber \\
x_N' &=& x_N + \frac{2 v_{\alpha}
\tau_0^\alpha}{\Gamma(\alpha+1)}(N+1/2)^\alpha \, .
\end{eqnarray}
Here, $v_{\alpha}=F_0/\eta_{\alpha}$ and $N$ counts the number of
time periods passed. The analytical solution~(\ref{xx}) for the mean
particle position $\langle x(t) \rangle$ from the
MFFPE~(\ref{FFPEmod})  is compared with the numerical solution of
the CTRW  in Fig.~\ref{Fig1} for different values of the fractional
exponent $\alpha$. The good agreement between our analytical and
numerical results confirms that Eq.~(\ref{FFPEmod}) is a correct
method to describe the CTRW driven by a rectangular time-dependent
force. Furthermore, the results depicted in Fig.~\ref{Fig1}
 exhibit the phenomenon of the ``death of linear response''
of the fractional kinetics to time-dependent fields in the limit $t
\to \infty$, reported also in Refs.~\cite{Barbi, SokolovKlafter06};
i.e. in the long-time limit the mean particle position approaches a
constant value, rather than being oscillatory, i.e.,
\begin{eqnarray} \label{asympt}
\langle x(\infty)\rangle=
v_{\alpha}\tau_0^{\alpha}b(\alpha)/\Gamma(\alpha+1),
\end{eqnarray}
where $b(\alpha)=\sum_{n=0}^{\infty}[2(n+1/2)^{\alpha}-n^{\alpha}-
(n+1)^{\alpha}]$, with the amplitude of the oscillations decaying to
zero as $1/t^{1-\alpha}$, see Eq.~(\ref{xav}). The function
$b(\alpha)$ describes the initial field phase effect which the
system remembers forever when $\alpha<1$. It changes monotonously
from $b(0)=1$ to $b(1)=0$. The averaged traveled distance $\langle
x(\infty)\rangle$ scales as $\tau_0^{\alpha}=$
$(2\pi/\Omega)^{\alpha}$, where $\Omega$ is the corresponding
angular frequency. This ``death of linear response'' to
time-periodic fields is also in agreement with the results for a
driven non-Markovian two state system \cite{PRL03} in the formal
limit of infinite  mean residence times.

%
\begin{figure}[ht]
\centering
\includegraphics[width=7.5cm]{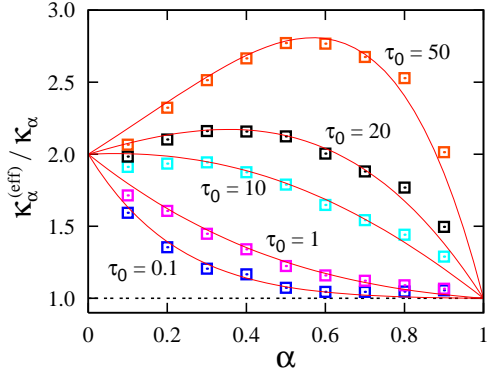}
\caption{(Color online) Scaled effective fractional diffusion
coefficient $\kappa_\alpha^{\mathrm{(eff)}}$ versus fractional
exponent $\alpha$ for different driving periods $\tau_0$. The
analytical prediction~(\ref{basic}) (continuous lines) is compared
with the numerical results (symbols) obtained from the CTRW by
averaging over $10^5$ trajectories. For $\tau_0>2\pi
\exp[-\frac{d}{d\alpha}\ln g(\alpha)|_{\alpha=0}] \approx 8.818$
the effective fractional diffusion coefficient
$\kappa_\alpha^{\mathrm{(eff)}}(\alpha)$ exhibits a maximum.}
\label{Fig2}
\end{figure}

We next study the mean square displacement and the effective
fractional diffusion coefficient $\kappa_\alpha^\mathrm{(eff)}$. We
recall that the free fractional diffusion is described by $\langle
\delta x^2(t)\rangle=2\kappa_{\alpha}
t^{\alpha}/\Gamma(1+\alpha)\propto t^{\alpha}$, while in the
presence of a constant bias, surprisingly, $\langle \delta
x^2(t)\rangle \propto t^{2\alpha}$
\cite{ScherMontroll,shlesinger1974}. Strikingly enough, the same
difference in the behavior remains true for a CTRW proceeding in a
periodic potential with zero bias for which $\langle \delta
x^2(t)\rangle \propto t^{\alpha}$ \cite{heinsalu2007} and in a
washboard potential with finite bias for which $\langle \delta
x^2(t)\rangle \propto t^{2\alpha}$ \cite{letter, heinsalu2006b},
described by the FFPE. The question thus arises, whether a
time-modulated subdiffusion follows the biased fractional
subdiffusion behavior $ \propto t^{2\alpha}$, or rather assumes the
unbiased behavior $ \propto t^{\alpha}$, as the average bias is
zero. To obtain the answer we use the Laplace-transform method and
the Fourier series expansion for the driving force
$F(t+\tau_0)=F(t)$ with frequency $\Omega=2\pi/\tau_0$,
\begin{eqnarray}
F(t)=\sum_{n=-\infty}^{\infty}f_n e^{in\Omega t}, \quad
f_{-n}=f_n^*
\end{eqnarray}
with $f_{2n}=0$ and $f_{2n+1}=-(2i/\pi)F_0/(2n+1)$ for the
rectangular driving under consideration. We assume that $\langle
x(0)\rangle=0$ and $\langle x^2(0)\rangle=0$ and denote the
 Laplace-transforms of the first and second moment by
$\tilde x(s)$ and $\tilde y(s)$, yielding
\begin{eqnarray}
s \tilde x(s)=\sum_{n=-\infty}^{\infty} v_n(s-in\Omega)^{-\alpha},
\end{eqnarray}
where $v_n=f_n/\eta_\alpha$. Because $\langle x(\infty)\rangle$ is
finite, we evaluate the asymptotical behavior of $\langle
x^2(t)\rangle$ rather than  $\langle \delta x^2(t)\rangle$.
Because the Laplace-transform of $D_t^{1-\alpha}\langle x(t)\rangle$
is $s^{-\alpha} s \tilde x(s)$,  the Laplace transform of the second
moment reads
\begin{eqnarray} \label{LT2nd}
s \tilde y(s)=2\kappa_\alpha/s^\alpha+2\sum_{m=-\infty}^{\infty}
v_m(s-im\Omega)^{-\alpha} \nonumber \\
\times \sum_{n=-\infty}^{\infty} v_n(s-i(m+n)\Omega)^{-\alpha}.
\end{eqnarray}
Note that in the double sum only the terms with $m=-n$ contribute
to the effective subdiffusion coefficient, being averaged over the
driving period. Therefore, we find that
\begin{eqnarray} \label{basic}
\kappa_\alpha^{\mathrm{(eff)}} & = & \kappa_\alpha+
2\frac{\cos(\pi\alpha/2)}{\Omega^\alpha} \sum_{n=1}^{\infty}\frac{
|v_n|^2}{n^\alpha} \nonumber \\
& = &
\kappa_\alpha+g(\alpha)F_0^2/(\eta_\alpha^2\Omega^\alpha),
\end{eqnarray}
where
\begin{eqnarray} \label{res2}
g(\alpha)=(2/\pi^2)\zeta(2+\alpha)[4-2^{-\alpha}]\cos(\pi\alpha/2)
\end{eqnarray}
is a   function decaying  from $g(0)=1$ towards $g(1)=0$ and
$\zeta(x)$ is the Riemann's zeta-function. From Eq.~(\ref{LT2nd})
one finds that the asymptotic behavior of the mean square
displacement is proportional to $t^{\alpha}$ as in the force free
case. It is now characterized, however, by an effective fractional
diffusion coefficient $\kappa_{\alpha}^\mathrm{(eff)}$ instead of
the free value $\kappa_\alpha$, i.e., $\langle \delta
x^2(t)\rangle=2\kappa_{\alpha}^\mathrm{(eff)}
t^{\alpha}/\Gamma(1+\alpha)$ for $t \to \infty$. The driving-induced
part of the effective subdiffusion coefficient is directly
proportional to the square of driving amplitude and inversely
proportional to $\Omega^\alpha$. For slowly oscillating force fields
this leads to a profound acceleration of subdiffusion as compared
with the force free case, see in Fig.~\ref{Fig2}: An optimal value
of the fractional exponent $\alpha$ exists, at which the
driving-induced part of the effective fractional diffusion
coefficient possesses a maximum.


In this letter we discussed the dynamics of anomalously slow
processes in time-varying potential landscapes within the CTRW and
FFPE descriptions. We demonstrated that the common form of the FFPE
given by Eq.~(\ref{FFPE}) is not valid for time-dependent forces; it
fails to correspond to the underlying CTRW modulated by an external
time-dependent force field. A modified form of the FFPE,
Eq.~(\ref{FFPEmod}), is derived for dichotomously alternating
force-fields. 
As an exactly solvable example we studied a periodic rectangular
force with zero average and successfully tested the analytical
results via numerical simulations of the underlying time-modulated
CTRW.

Our study, however, is not able to validate the correctness of the
MFFPE~(\ref{FFPEmod}) when  extended {\it ad hoc} to an arbitrary
time-dependent potential landscape different from the dichotomous
case.
A description of  time-dependent fields via subordination in
conjunction with a CTRW approach is also doomed to failure  because
of the distinct difference between the deterministic physical time
and the merely mathematical random subordination time. As a matter
of fact, any slowly non-zero frequency time-varying force varies
infinitely fast within the realm of fictitious, operational
subordination time. This causes CTRW subdiffusion  to fail in
responding to time-periodic fields in an ordinary manner. In
addition, all those theories modeling dielectric response which are
based on such an approach are thus also physically defeasible. A way
out of this dilemma consists in relying on models of driven
subdiffusion which either are based on  the generalized Langevin
dynamics \cite{Kubo} or  on fractal Brownian motion. The challenge
of modeling  subdiffusion   in a time-varying potential landscape
thus necessitates plenty of further enlightening research.


We thank E.~Barkai, R.~Metzler, and I.~Sokolov for constructive
 discussions. This work has been supported by the
Estonian Science Foundation via grant no. 6789,  the Archimedes
Foundation (EH),  the DFG-SFB-486 (PH), and by the Volks\-wagen
Foundation, no. I/80424 (PH).


\end{document}